\documentstyle[twocolumn,aps,prb,epsf]{revtex}
\begin{document}
\title{
Pinned Balseiro-Falicov Model of Tunneling and Photoemission in the Cuprates}

\author{R.S. Markiewicz$^{1,2}$, C. Kusko$^{1,2,*}$, and V. Kidambi$^1$} 

\address{Physics Department (1) and Barnett Institute (2), 
Northeastern U.,
Boston MA 02115}
\maketitle
\begin{abstract}
The smooth evolution of the tunneling gap of Bi$_2$Sr$_2$CaCu$_2$O$_8$ 
with doping from a pseudogap state in the underdoped cuprates to a
superconducting state at optimal and overdoping, has been interpreted as
evidence that the pseudogap must be due to
precursor pairing.  We suggest an alternative explanation,
that the smoothness reflects a hidden SO(N) symmetry near the $(\pi ,0)$ points
of the Brillouin zone (with N = 3, 4, 5, or 6).  Because of this symmetry, the
pseudogap could actually be due to any of a number of nesting instabilities,
including charge or spin density waves or more exotic phases.
\par
We present a detailed analysis of this competition for one particular model:
the pinned Balseiro-Falicov model of competing charge density wave and (s-wave)
superconductivity.  We show that most of the anomalous features of both 
tunneling and photoemission follow naturally from the model, including the
smooth crossover, the general shape of the pseudogap phase diagram, the 
shrinking Fermi surface of the pseudogap phase, and the asymmetry of the 
tunneling gap away from optimal doping.  Below $T_c$, the sharp peak at $\Delta
_1$ and the dip seen in the tunneling and photoemission near $2\Delta_1$ cannot 
be described in detail by this model, but we suggest a simple generalization to
account for inhomogeneity, which does provide an adequate description.

We show that it should be possible, with a combination of photoemission and
tunneling, to demonstrate the extent of pinning of the Fermi level to the
Van Hove singularity.  A preliminary analysis of the data suggests pinning in
the underdoped, but not in the overdoped regime. 
\end{abstract}


\narrowtext

\section{Introduction}

\subsection{Precursor Pairing?}

Recent photoemission\cite{Gp0,Gp1,Gp2} and 
tunneling\cite{tu4,tu6,tu1,tu2,tu3,tu5} studies have provided a picture of 
unparalleled detail of the opening of the pseudogap in the underdoped 
cuprates.  The most remarkable feature is that the pseudogap evolves smoothly
into the superconducting gap as doping increases.  This has led a number of 
researchers to conclude that the pseudogap must itself be related to
superconductivity: that it represents a form of short-range superconducting
order, or precursor pairing.  It is the purpose of the present paper to show
that this is not a foregone conclusion: there is an alternative interpretation
(better: class of interpretations) in which the pseudogap represents a {\it
competing} ordered state.  In this case, the apparently smooth evolution is due 
to an {\it underlying symmetry of the instabilities} of the problem -- a 
manifestation of an SO(N) group, with N = 3\cite{SO3}, 4\cite{Zhang4}, 
5\cite{Zhang5}, or 6\cite{SO6}.
\par
For such an interpretation to hold, certain very strict conditions must be met.
Specifically, there must be a Van Hove nesting\cite{RiSc}, with the Van Hove 
singularity (VHS) pinned close to the Fermi level $\epsilon_F$ over an extended 
doping range.  This prediction is now within the realm of experimental test, and
a preliminary analysis of the existing data seems to confirm the pinning.
\par
To describe the competing order parameters, we analyze a simple generic model, 
the pinned Balseiro-Falicov (BF)\cite{BFal} model.  Within this model, the total
gap has a maximum at $(\pi ,0)$ in the Brillouin zone, given by 
\begin{equation}
\Delta_t=\sqrt{\Delta_k^2+G_k^2}, 
\label{eq:1c}
\end{equation}
where $\Delta_k$ is an (s-wave) superconducting gap and
$G_k$ a charge-density wave (CDW) gap, defined below.  This is exactly the form 
proposed phenomenologically by Loram, et al.\cite{Lor}, and it immediately
explains the smooth evolution of the gap with doping: there is a single gap
(at $(\pi ,0)$) at all dopings, even though the system changes over from
CDW near half filling to superconductor near optimal doping.  The form of 
Eq.\ref{eq:1c} immediately follows from an approximate SO(4) symmetry.  It is
in principle possible to disentangle the nature of the gaps from their behavior
away from $(\pi ,0)$, although this involves symmetry breaking terms, and hence
is considerably more model dependent.  The SO(N) symmetry also suggests that the
pinned BF (pBF) model should provide a good approximate representation for a
wide range of competing phases -- in particular the pseudogap phase may also be
an antiferromagnetic phase or a flux phase (which is a form of dynamic CDW),
or indeed a striped phase which is a combination of two of these phases.

\subsection{Van Hove Pinning}

In the generalized Van Hove scenario\cite{Surv}, there are two separate 
phenomena which contribute to Van Hove pinning.  First, as part of the 
Mott-Hubbard transition, strong Hubbard-U correlation effects
renormalize Cu-O hopping $t$ to zero at half filling, leaving a residual energy
dispersion associated with exchange $J$.  In the absence of correlation effects,
the `bare' VHS would fall at a finite hole doping fixed by the
band parameters ($t^{\prime}$ in the $tt^{\prime}J$ model) -- this doping will
be close to, but not necessarily the same as optimal doping.  However, since 
second (Cu) neighbor exchange is expected to be small, the exchange bands have 
the simple dispersion $J(c_x+c_y)$, where $c_i=\cos{k_ia}$.  Hence, at half 
filling, the Fermi level approximately coincides with the VHS.
This phase can be further stabilized by Van Hove nesting, which opens up a large
gap in the dispersion near $(\pi ,0)$.  A good fit to the dispersion in the 
related insulating compound\cite{Well} Sr$_2$CuO$_2$Cl$_2$ (SCOC) can be 
found\cite{Laugh,WeL,Pstr} by assuming that the nesting is associated with a
flux phase\cite{Affl}.

With doping, $t$ is gradually restored, producing a behavior which cannot be 
described by a rigid band filling model.  At first, the VHS shifts faster than 
the Fermi level, so the VHS lies {\it below} the Fermi level, but very close to 
it.  Gradually, $t$ saturates, the VHS stops shifting (usually close to its bare
value), and at some finite doping $x_c$ the Fermi level crosses this new VHS.  
Thus, the Fermi level coincides with the VHS {\it twice}: at half filling and 
near the bare VHS, and remains anomalously close at intermediate dopings.  
Since its initial discovery\cite{RM3}, this correlation induced pinning has been
confirmed by a number of calculations\cite{Surv1}.  

But even stronger pinning is possible.  Near $x_c$, the energy can be 
significantly lowered by a second nesting instability.  Assuming this second
instability to be charge-density wave (CDW) related, a self-consistent 3-band
slave boson calculation demonstrated that this model can lead to two free
energy minima, one at half filling and the other at $x_c$\cite{Pstr}. 
This results in striped phases, with each phase pinned near a VHS:
the magnetic stripes near the J-dominated VHS near half filling, and the
charged stripes near the t-dominated VHS at $x_c$.  Since the stripes are
nanoscale (due to long-ranged Coulomb repulsion), the system evolves rather
smoothly with doping, with the Fermi level remaining even more strongly pinned 
to the VHS at all dopings.

While the above model is in good agreement with experiments on both the
pseudogaps and the striped phases\cite{Tran}, the identification of the
specific nesting instability phases is less secure.  This is because the
nesting and pairing instabilities of the Van Hove scenario form a group, SO(6),
and the instabilities associated with two dual 6-dimensional superspins are
nearly degenerate\cite{SO6}, so the issue of which phase is the most unstably
depends sensitively upon secondary parameters which are not well known.  The
various possibilities include the antiferromagnetic and d-wave superconducting
instabilities of Zhang's SO(5)\cite{Zhang5}, as well as CDW's, flux phases, and
a more exotic spin current phase\cite{Sch2}.  (Two-leg ladders have an even 
larger assortment of instabilities to choose from\cite{SO8}.)  It is difficult
to incorporate the details of this phase separated regime rigorously into
calculations of the ARPES and tunneling spectra.

Nevertheless, a remarkably simple picture of the pseudogap (VH nesting) phase
diagram can be developed via a simple {\it Ansatz} for the pinned striped 
phases.  A one band model is assumed
\begin{equation}
\epsilon_k=-2t(c_x+c_y)-4t^{\prime}c_xc_y
\label{eq:1b}
\end{equation}
(with $c_i=\cos{k_ia}$), and the second neighbor hopping is adjusted with doping
to pin the VHS to the Fermi level over a doping range from half filling $x=0$ to
a doping $x_c$.  Approximately\cite{Surv},
\begin{equation}
\tau={2t^{\prime}\over t}=-1.04\tanh{2.4x}.
\label{eq:2}
\end{equation}
Within this model the striped phase is represented by a single nesting 
instability which splits the band dispersion at $(\pi ,0)$.  Such a model was
initially introduced (without the pinning) by Balseiro and Falicov\cite{BFal} to
study the competition between CDW's and s-wave superconductivity.  We have
employed this pinned Balseiro-Falicov (pBF) model in previous pseudogap 
studies\cite{RM8a,RMPRL,MKu} and will
continue to use it here.  [It must be stressed that we do not employ the CDW's
to mimic the spatial pattern of the stripes, but rather to reproduce the Fermi
level pinning.]  We are currently generalizing the model to include d-wave
superconductivity and a variety of other nesting instabilities.  In the present
version, there are electron-phonon coupling energies $\lambda_G$ associated with
CDW's, and $\lambda_{\Delta}$ with superconductivity, which may or may not be 
equal.  The gap equations are solved self-consistently, with free parameters 
$t$, $x_c$, the $\lambda_i$'s, and $\omega_{ph}$, a bosonic cutoff 
frequency.  There is relatively little data in the overdoped regime, so we 
choose a simple model: for $x>x_c$, the band parameters cease evolving, and the
additional holes just rigidly fill the band, shifting the Fermi level away from 
the VHS.  This simple picture appears capable of describing the overdoped
regime in both YBa$_2$Cu$_3$O$_{7-\delta}$ (YBCO), where the pseudogap vanishes
close to optimal doping\cite{Gp3} and La$_{2-x}$Sr$_x$CuO$_4$ (LSCO), where the
pseudogap appears to persist well into the overdoped regime\cite{Gp6}.
However, a more complicated behavior, including a second range of two-phase 
coexistence (Sections 11.6-11.7 of Ref. \cite{Surv}), is not ruled out.

Within this picture, there is a natural doping dependence associated with the
competition between nesting and pairing instabilities.  At half filling, 
$t^{\prime}=0$, the perfect nesting overwhelms the pairing instability, leading 
to a pure nesting instability.  As $t^{\prime}$ increases with doping (to 
maintain the VH pinning), the nesting gets worse, while pairing is less affected
(it is actually enhanced, Section II.B).  This leads to a crossover to a pairing
instability as a function of doping.  Since nesting does not gap the full Fermi 
surface (when $t^{\prime}$ is large enough), superconductivity can appear at a 
lower temperature, with a $T_c$ which
increases with doping.  As long as the VHS remains pinned to the Fermi level
and the electron-phonon coupling remains doping independent, the nesting $T_n$
will decrease with doping $x$ while the superconducting $T_c$ will increase.
At some point, the two transitions would cross.  However, at this point $T_n$
will be rapidly suppressed to zero, since superconductivity gaps essentially the
full Fermi surface (the small residual Fermi surface for a d-wave gap could only
sustain a nesting instability at a much lower $T$).  While exactly this
behavior is found in YBCO, in LSCO and probably Bi$_2$Sr$_2$CaCu$_2$O$_8$ 
(Bi-2212) as well the pseudogap persists into the overdoped regime.  This 
behavior can be modelled by having the strength of the bosonic pairing decrease 
with increasing $x$.

Since nesting splits the VHS degeneracy, we introduce some notation to clarify
the following discussions.  The {\it lower VHS} (VHL) is the VHS shifted below
the Fermi level, at energy $E_L$, and hence visible in photoemission. The {\it
upper VHS} (VHU) is shifted to $E_U$, above the Fermi level, and hence can
only be seen in tunneling (or inverse photoemission).  In conventional plots of
tunneling spectra, VHL appears at negative voltages, in the electron extraction
mode, and VHU at positive voltages.  The {\it Van Hove centroid} (VHC) is the
average position of these two features: $E_C=(E_U+E_L)/2$ -- it is the position
at which the single, unsplit VHS would fall at high temperatures, well above
the pseudogap transition.  

In terms of these features, the photoemission gap is $\Delta_{PE}=\epsilon_F-
E_L$.  We will show (Fig.~\ref{fig:3} below) that the tunneling gap is 
$\Delta_{TU}=(E_U-E_L)/2$.  In this case, the pinning of the Fermi level at the 
VHC, $\epsilon_F\simeq E_C$, can be rewritten in terms of measurable quantities 
as $\Delta_{TU}=\Delta_{PE}$.  In Section III.C, below, we will demonstrate that
this relationship appears to be satisfied in the underdoped regime.

\section{Striped Phases and Pseudogap}

\subsection{Phase Diagram of Pinned BF Model}

For completeness, we recall the energy dispersion and the gap equations of the 
BF model.  In terms of the four-component wave vector $\Psi_{\vec k}^{\dagger}
=(\psi_{\vec k}^{\dagger},\ \psi_{\vec k+\vec Q}^{\dagger},\ \psi_{-\vec k},\ 
\psi_{-(\vec k+\vec Q)})$, the mean field BF Hamiltonian tensor is
\begin{equation}
H_{BF}=\left(\matrix{\epsilon_{\vec k}-\epsilon_F&-G_{\vec k}&-\Delta_{\vec k}&
     0\cr
   -G_{\vec k}&\epsilon_{\vec k+\vec Q}-\epsilon_F&0&-\Delta_{\vec k+\vec Q}\cr
   -\Delta_{\vec k}&0&-\epsilon_{\vec k}+\epsilon_F&G_{\vec k}\cr
   0&-\Delta_{\vec k+\vec Q}&G_{\vec k}&-\epsilon_{\vec k+\vec q}+\epsilon_F}
     \right).
\label{eq:21}
\end{equation}
In terms of a function
\begin{equation}
\Theta_{\vec k}=\cases{1,&if $|\epsilon_{\vec k}-\epsilon_F|<\hbar\omega_{ph}$;
                      \cr
                       0,&otherwise,\cr}
\label{eq:22}
\end{equation}
the gap functions are $\Delta_{\vec k}=\Delta\Theta_{\vec k}$ for 
superconductivity, and $G_{\vec k}=G_0+G_1\Theta_{\vec k}\Theta_{\vec k+\vec Q}$
for the CDW.  The energy eigenvalues are $E_{\pm,k}$ and their negatives, with
\begin{equation}
E_{\pm,k}^2={1\over 2}(E_k^2+E_{k+Q}^2+2G_k^2\pm\hat E_k^2),
\label{eq:3}
\end{equation}
$E_k^2=\epsilon_k^2+\Delta_k^2$, $\hat E_k^4=(E_k^2-E_{k+Q}^2)^2+4G_k^2\tilde 
E_k^2$, $\tilde E_k^2=\epsilon_{k+}^2+(\Delta_k-\Delta_{k+Q})^2$,
$\epsilon_{k\pm}=\epsilon_k\pm\epsilon_
{k+Q}$, and the nesting vector $Q=(\pi ,\pi )$. 
If the magnitudes of the (attractive) phonon-induced electron-electron 
interaction energies are $\lambda_{\Delta}$ and $\lambda_G$, then the gap 
equations are
\begin{eqnarray}
\Delta=\lambda_{\Delta}\Delta\Sigma_{\vec k}\Theta_{\vec k}{1\over E_{+,k}^2-
E_{-,k}^2}\Bigl({E_{+,k}-E_{-,k}\over 2}-
\nonumber \\
({1\over 2E_{+,k}}-{1\over 2E_{-,k}})(\epsilon_
{\vec k+\vec Q}^2+\Theta_{\vec k+\vec Q}[\Delta^2+G_{\vec k}^2]\Bigr),
\label{eq:23}
\end{eqnarray}
\begin{eqnarray}
G_i=\lambda_G\Sigma_{\vec k}\Theta_i{G_{\vec k}\over E_{+,k}^2-E_{-,k}^2}
\Bigl({E_{+,k}^2+\epsilon_{\vec k}\epsilon_{\vec k+\vec Q}-\Delta_{\vec k}^2-
G_{\vec k}^2\over 2E_{+,k}}-
\nonumber \\
{E_{-,k}^2+\epsilon_{\vec k}\epsilon_{\vec k+\vec Q}-\Delta_{\vec k}^2-
G_{\vec k}^2\over 2E_{-,k}}\Bigr),
\label{eq:24}
\end{eqnarray}
with $\Theta_0=\Theta_{\vec k}\Theta_{\vec k+\vec Q}$, $\Theta_1=1$.  A similar
model\cite{KOP} has recently been applied to analyze the photoemission
associated with a pure CDW phase.

In the pinned BF (pBF) model, the Fermi level is pinned to the VHC, via 
Eq.~\ref{eq:2}, for doping $x$ between half filling ($x=0$) and some critical
doping $x_c$, while for $x>x_c$, the curvature is fixed at $t^{\prime}=t^
{\prime}(x_c)$, while the Fermi level shifts off of the VHC in a rigid band
filling.
\begin{figure}
\leavevmode
   \epsfxsize=0.33\textwidth\epsfbox{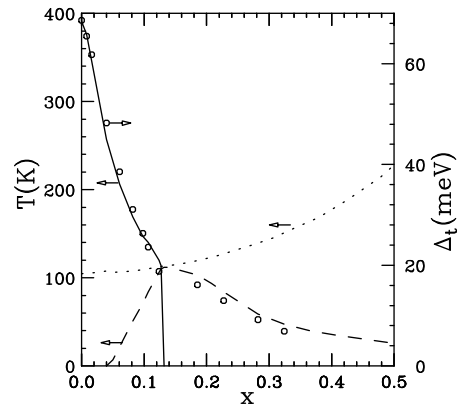}
\vskip0.5cm 
\caption{Phase diagram of pinned Balseiro-Falicov model. Circles = net tunneling
gap, $\Delta_t$; solid line = CDW transition temperature $T_p$; dashed line =
superconducting transition T$_c$; dotted line = $T_c$ in absence of CDW, $x_c>
0.5$.  }
\label{fig:1}
\end{figure}

\begin{figure}
\leavevmode
   \epsfxsize=0.33\textwidth\epsfbox{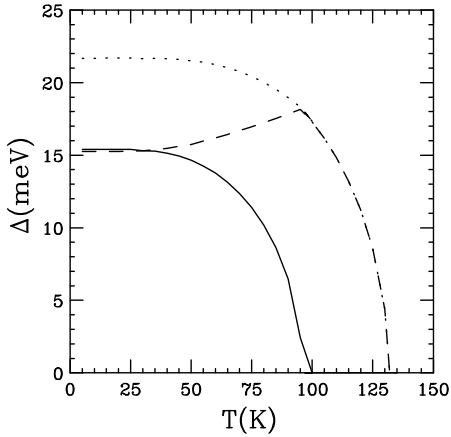}
\vskip0.5cm 
\caption{Evolution of gaps in pinned Balseiro-Falicov model. Solid line =
superconducting gap $\Delta_k$; dashed line = CDW gap $G_k=G_0+G_1$; dotted line
= net gap, $\Delta_t$, Eq.~\protect\ref{eq:1}.  Parameters correspond to 
Fig~\protect\ref{fig:1}, $x=0.1$.  }
\label{fig:1d}
\end{figure}

For fixed values of the parameters $t$, $x_c$, $\lambda_i$, and phonon cutoff 
$\omega_{ph}$, the pseudogap phase diagrams are derived by solving 
Eqs.~\ref{eq:3}-\ref{eq:24} self-consistently.  For Fig.~\ref{fig:1}, the 
parameter values are chosen as $t=\lambda_{\Delta}=\lambda_G=0.25eV$, 
$\omega_{ph}=45meV$, and $x_c=0.123$.  At half filling, 
perfect nesting wins out over superconductivity, but with doping the nesting
becomes poorer, while superconductivity is enhanced, leading to a crossover.
As soon as the superconducting $T_c$ is larger than $T_{CDW}$, $T_{CDW}$ is 
rapidly suppressed to zero.  The dotted line in Fig.~\ref{fig:1} shows how $T_c$
would evolve in the absence of the CDW, if the Fermi level remained pinned to 
the VHC over the full doping range ($x_c>0.5$).  A typical temperature 
dependence of the resulting gaps is illustrated in Fig.~\ref{fig:1d}, for $x=
0.1$.  Note that $\Delta_t$, Eq.~\ref{eq:1}, is found self-consistently to 
evolve smoothly with temperature; this implies that $G_k$ must actually decrease
when superconductivity appears, $\Delta_k\ne 0$.

\par
\begin{figure}
\leavevmode
   \epsfxsize=0.33\textwidth\epsfbox{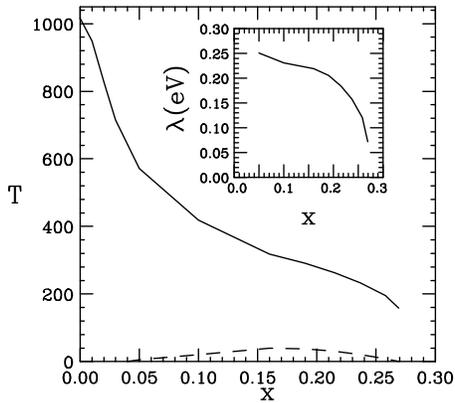}
\vskip0.5cm 
\caption{Model pseudogap phase diagram for LSCO.  Parameters $x_c=0.16$, 
$\lambda_G=0.6eV$, with doping dependent $\lambda_{\Delta}$ (inset).  Solid line
 = CDW transition temperature $T_d$; dashed line = superconducting transition 
T$_c$.}
\label{fig:21}
\end{figure}

To reproduce the phase diagram of YBCO, the crossover must arise close to $x_c$,
as in Fig.~\ref{fig:1}, while in LSCO it falls well after $x_c$, Fig. 
\ref{fig:21}.  The phase diagram of LSCO can be modelled by choosing 
$\lambda_G\ne\lambda_{\Delta}$, and 
letting the latter vary with doping, Fig.~\ref{fig:21}.
\par
One feature of Fig. \ref{fig:21} should be noted.  In LSCO the pseudogap appears
in some experiments to persist into the overdoped regime\cite{Gp6,Gp8}, while 
other experiments find a striking crossover in properties at optimal 
doping\cite{Surv2,Boeb,Naei}.  In the thermodynamic experiments\cite{Gp8}, the
pseudogap appears to be closing linearly with doping, but then has a {\it break
in slope at optimal doping}, and falls off more slowly at larger doping.  Figure
\ref{fig:21} mimics this behavior: the Fermi level is pinned at the VHC until
optimal doping $x_c=0.16$, and then depins.  Despite the fact that the Fermi 
level now shifts with doping away from the VHC, the CDW instability is strong
enough that the CDW phase persists out to $x=0.27$, at which point it becomes 
unstable (discontinuously).  Such behavior is only possible because the
superconducting transition is suppressed far below $T_p$.

The pBF model is a 2D mean field theory.  In a more accurate three-dimensional
calculation\cite{RM5}, the transition temperatures in the above phase diagram
are likely to be replaced by crossover temperatures at which 2D fluctuations
become strong -- i.e., temperatures at which a pseudogap opens.  When interlayer
coupling becomes strong enough, a real three-dimensional order, with true
energy gaps, can develop.  

A (weak-coupling) limitation of the BF model is the discontinuous change in
$\Delta$ and $G_1$ when the energy crosses $\omega_{ph}$.  When this feature
becomes prominent in the dispersion, the BF results must be considered as only a
qualitative indication of the results of a proper strong coupling calculation.

\subsection{SO(6)}

Whereas the above discussion has been in terms of a particular competition --
between a CDW and an s-wave superconductor, the results are indicative of a
much more general situation.  That is because the VHS instability has an 
underlying SO(6) symmetry group\cite{SO6}.  There is actually a pair of 
6-dimensional `superspins', which consist of various nesting or pairing 
instabilities of the VHS.  Nesting (pairing) operators are those instability 
operators which do (do not) commute with the number operator Q.
\par
For example, Zhang's SO(5) group\cite{Zhang5} is a 
subgroup of this SO(6), where the
competition is between antiferromagnetism (or a spin-density wave -- SDW) and
d-wave superconductivity.  The SO(6) offers greater flexibility, with the
nesting instability being either the CDW or SDW, or a flux phase\cite{Affl} or
a spin current instability\cite{Sch2}.  

\begin{figure}
\leavevmode
   \epsfxsize=0.33\textwidth\epsfbox{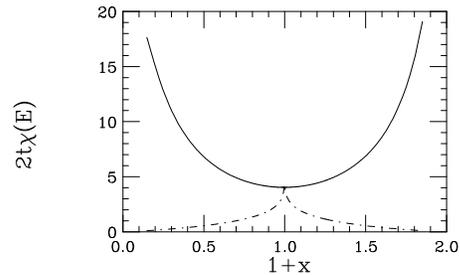}
\vskip0.5cm 
\caption{Susceptibilities $\chi_{Q}$ (dotdashed line) and $\chi_{0}$ (solid
line) vs. band filling $1+x$ for Eq.~\protect\ref{eq:1b} with $4t^{\prime}=
E_F$.}
\label{fig:4}
\end{figure}

The generic features of the instabilities can be
most clearly seen be looking at the doping dependence of the corresponding 
(bare) susceptibilities $\chi_0(\vec q,\omega)$.  For a pinned VHC, this is 
illustrated in Fig.~\ref{fig:4}\cite{SO6}.  For all the instabilities, there are
only two bare susceptibilities, $\chi_0=\chi_0(\vec q=0,0)$ for the pairing 
instabilities and $\chi_Q=\chi_0(\vec q=\vec Q,0)$ for the nesting 
instabilities.  At half filling these susceptibilities are degenerate, but as 
$t^{\prime}$ increases, the VHS's become more one-dimensional as the planes 
become more like interpenetrating Cu-O-Cu chains\cite{RM8b}.  In this case, 
nesting is greatly reduced,
since the chains run at right angles, whereas pairing is enhanced, because the
1D bands have a stronger VHS.  
\par
We note that the phase diagram of Fig.~\ref{fig:1} shows considerable evidence
for the underlying SO(N) nature of the VHS: (a) despite a crossover from CDW to
superconducting, the evolution of the total gap $\Delta_t$ and transition
temperature $T_t$ with doping is extremely smooth; indeed, the gap ratio 
$2\Delta (0)/k_BT_c\simeq 4.1$ is nearly independent of doping.  (b) The total
gap is given by a vector addition, Eq.~\ref{eq:1c}. And (c)
the overall shapes of the individual (decoupled) transitions $T_i(x)$ resemble
the doping dependences of the generic susceptibilities, Fig.~\ref{fig:4}.
Thus, the solid line in Fig.~\ref{fig:1} (pseudogap phase)
closely resembles the nesting susceptibility (a larger susceptibility 
corresponding to a higher transition temperature), while the pairing 
susceptibility resembles the superconducting phase (dotted lines in 
Fig.~\ref{fig:1}).  
\par
Note finally that just these
features -- in particular feature (a) -- have been taken as evidence that the
pseudogap must itself be related to superconductivity as a form of precursor 
pairing.  Instead, we find that the phase diagram of Fig.~\ref{fig:1}
should look fairly similar for {\it any} competition between a nesting and a
pairing instability -- e.g., either CDW vs s-wave superconductivity or SDW vs
d-wave superconductivity.  The actual phases observed will depend on the
interaction parameters, and may vary with doping.  
Hence, it seems unlikely that the experimental 
pseudogap represents superconducting fluctuations, unless all nesting 
instabilities are somehow strongly suppressed.
\par
We stress that the similarity between Figs. \ref{fig:1} and \ref{fig:4}
exists {\it even though the BF model does not possess SO(4) symmetry}.
The form of the phonon coupling leads to very different gap equations for
CDW and superconductivity, Eqs. \ref{eq:22} - \ref{eq:24}.  Thus, the two
instabilities are not degenerate at zero doping.  Yet the doping dependence of 
the two instabilities follows the susceptibilities of Fig. \ref{fig:4}, and the
gap at $(\pi ,0)$ is just the total `length' of the two individual gaps (adding
in quadrature).

Experimentally, there is evidence for striped phases\cite{Tran} at intermediate 
dopings.  In SO(5), these stripes can be interpreted as a combination of 
magnetic and superconducting stripes.  However, in this case it is not clear how
to interpret the 1/8 anomaly ($x=1/8$) where the stripes show long range order 
but there is no superconductivity.  It seems more likely that superconductivity
competes with the stripes, and that the stripes arise from a competition
between two nesting instabilities, the flux phase at half filling and a CDW near
$x_c$\cite{Pstr}.  It is important to recognize that the flux phase does not 
involve the electron spins -- the magnetic moments are orbital.  Hence, the 
phase is best understood as a {\it dynamic CDW phase} (compare 
Ref.~\cite{RM8c}), so a model which approximates the pseudogap by a CDW {\it 
Ansatz} (such as the pBF model) should provide a quite reasonable first 
approximation.

\subsection{Tunneling Spectra}

Under certain special conditions, both the tunneling and photoemission can
reveal very direct information about the spectral function of the interacting 
electrons.  The quasiparticle tunneling current can be written\cite{Mah}
\begin{eqnarray}
I=2e\Sigma_{\vec k\vec p}|T_{\vec k\vec p}|^2\int_{-\infty}^{\infty}{d\epsilon
\over 2\pi}A_R(\vec k,\epsilon)A_L(\vec p,\epsilon +eV)\times
\nonumber \\
{[n_F(\epsilon )-n_F(\epsilon +eV)]},
\label{eq:31}
\end{eqnarray}
where $T_{\vec k\vec p}$ is the tunneling matrix element, $A_i$ is the 
appropriate spectral function in the metal on the left ($L$) and right ($R$) of
the tunneling junction.  If the tunneling matrix element is considered to be
constant, independent of $\vec k$ and $\vec p$, it can be taken out of the
integral, yielding
\begin{equation}
I=2eT^2\int_{-\infty}^{\infty}{d\epsilon\over 2\pi}N_R(\epsilon)N_L(\epsilon 
+eV)[n_F(\epsilon )-n_F(\epsilon +eV)],
\label{eq:42}
\end{equation}
with $N_i$ the appropriate dos.  For an NIS junction, taking $N_L$ to be energy
independent, the tunneling conductance is
\begin{equation}
G={\partial I\over\partial V}={e^2T^2\over\pi}N_LN_R(-eV).
\label{eq:33}
\end{equation}

The conventional wisdom\cite{WAH}
is that tunneling is not sensitive to the dos: the explicit factor of dos in
Eq. \ref{eq:31} is cancelled by $T^2\propto v_F,$ the Fermi velocity.  However, 
Wei, et al.\cite{Wei} have shown that this cancellation breaks down in the 
presence of strong anisotropy.  In particular, {\it tunneling along the 
c-direction into a two-dimensional metal directly measures the in-plane dos} (at
least in the thin junction limit).  Hence, in accord with Eq.~\ref{eq:33}, the 
tunneling conductance is proportional to the tunneling dos.  This result is the 
basis for the present analysis.  For three-dimensional materials, there have
been previous proposals that the VHS's should show up in tunneling\cite{VaPF},
and the tunneling spectra of CDW superconductors have been analyzed\cite{Gab}.

The tunneling and photoemission are derived from the spectral function of the
model
\begin{equation}
A(k,\omega )=2\pi\Sigma_{i=\pm} [u_{i,k}^2\delta(\omega -E_{i,k})+v_{i,k}^2
\delta(\omega +E_{i,k})],
\label{eq:1}
\end{equation}
with eigenenergies given by Eq.~\ref{eq:3}, and coherence factors
\begin{equation}
u_{\pm,k}=u_{\pm,0}\cos({\phi +\phi_{\mp}}),
\label{eq:2a}
\end{equation}
$v_{\pm,k}^2=u_{\pm,0}^2-u_{\pm,k}^2$, 
\begin{equation}
u_{+,0}^2=1-u_{-,0}^2={1\over 2}(1+{E_k^2-E_{k+Q}^2\over\hat E_k^2}),
\label{eq:2b}
\end{equation}
\begin{equation}
\cos^2\phi ={1\over 2}(1+{\epsilon_k+\epsilon_{k+Q}\over\tilde E_k}),
\label{eq:2c}
\end{equation}
\begin{equation}
\cos^2{\phi_{\pm}} ={1\over 2}(1+[{E_{\pm,k}^2+\epsilon_k
\epsilon_{k+Q}-\Delta_k\Delta_{k+Q}+G_k^2\over \tilde E_kE_{\pm,k}}]),
\label{eq:2d}
\end{equation}
\begin{equation}
\sin{2\phi} ={\Delta_k-\Delta_{k+Q}\over\tilde E_k},
\label{eq:3c}
\end{equation}
\begin{equation}
\sin{2\phi_{\pm}} ={\Delta_k\epsilon_{k+Q}+\Delta_{k+Q}\epsilon_k\over E_{\pm,k}
\tilde E_k}.
\label{eq:3d}
\end{equation}
In deriving the spectral function, it is convenient to use the group
theoretical techniques of spectrum-generating algebras\cite{SO6}.

For a pure CDW ($\Delta_k=0$), the spectral function simplifies to 
\begin{equation}
A(k,\omega )=2\pi [u_k^2\delta(\omega -E_{k+})+v_k^2\delta(\omega -E_{k-})],
\label{eq:11}
\end{equation}
with
$u_k^2=1-v_k^2=(1+\epsilon_{k-}/ \tilde E^{\prime}_k)/2$,
\begin{equation}
E_{k\pm}=(\epsilon_{k+}\pm\tilde E^{\prime}_k)/2,
\label{eq:11a}
\end{equation}
and $\tilde E^{\prime}_k=\sqrt{\epsilon_{k-}^2+4G_k^2}$.

\begin{figure}
\leavevmode
   \epsfxsize=0.33\textwidth\epsfbox{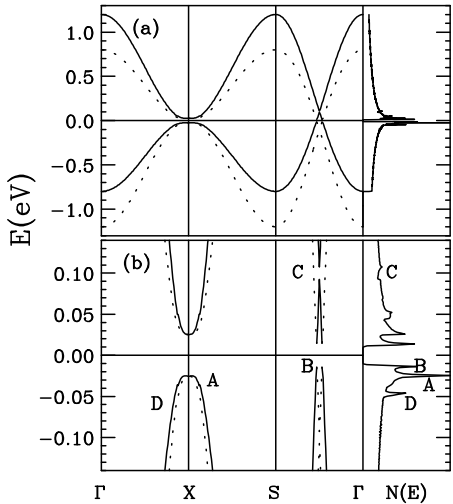}
\vskip0.5cm 
\caption{Energy dispersion for data of Fig.~\protect\ref{fig:1}, with $\tau =
-0.2$, $T=10K$. Right = tunneling dos. (a) = full dispersion; (b) = blow-up of 
region near $\epsilon_F$.}
\label{fig:3}
\end{figure}

\begin{figure}
\leavevmode
   \epsfxsize=0.33\textwidth\epsfbox{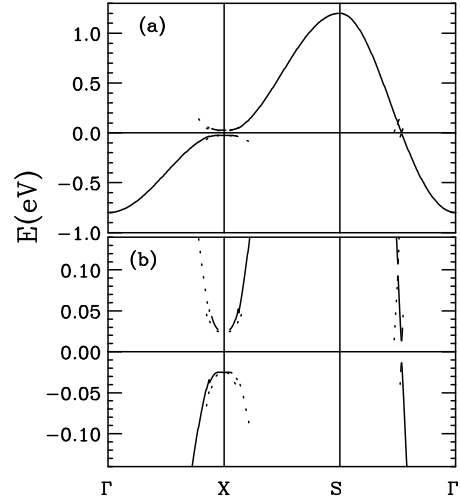}
\vskip0.5cm 
\caption{Replot of energy dispersion of Fig.~\protect\ref{fig:3}, illustrating
spectral weight.  Coherence factor $\ge 0.6$: solid lines; between 0.1 and 0.6:
dashed lines; between 0.001 and 0.1: dotted lines.}
\label{fig:3a}
\end{figure}

Figure~\ref{fig:3} shows the energy dispersion for a combined CDW-superconductor
(left) and the associated dos (right).  Part (a) shows the full dispersion, 
while (b) is a blow-up of the region near the Fermi level.  There are four 
bands; the CDW order folds the M-point of the Brillouin zone into the $\Gamma$
point, while superconductivity folds both of these bands around the Fermi level
(dotted lines).  However, these ghost bands carry little spectral weight
(coherence factor $<<1$) except very near the Fermi level -- note that there is
no indication of the tops or bottoms of the dotted lines in the dos.  
Figure~\ref{fig:3a} replots the dispersion, giving an indication of the spectral
weight.  See further Fig.~\ref{fig:11}, below.

A close look at the region near the Fermi level, Fig.~\ref{fig:3}b, reveals
that structure in the tunneling dos is directly related to
features in the dispersion of the gapped bands.  Thus, peak A is associated with
the dispersion at $(\pi ,0)$ -- the VHS peak split by the combined
CDW-superconducting gap.  Peak B is due to the superconducting gap away from
$(\pi ,0)$ -- particularly near $(\pi /2,\pi /2)$.  Accordingly, it will be
considerably less prominent for a d-wave superconductor, where the gap vanishes
at $(\pi /2,\pi /2)$.  Feature C is associated with the CDW gap $G_k$ near
$(\pi /2,\pi /2)$. 
As discussed in Ref.~\cite{Pstr}, Equation~\ref{eq:11a} can be rewritten as 
\begin{equation}
E_{k\pm}=-4t^{\prime}c_xc_y\pm\sqrt{4t^2(c_x+c_y)^2+G_k^2},
\label{eq:11b}
\end{equation}
showing that the CDW gap is fixed to the VHS, but not to the Fermi level, so
that at $\vec k$-points away from $(\pi ,0)$ there can be two gaps at different
energies.  Feature C is further discussed in Ref.~\cite{KOP}. 
Whether this is only a weak-coupling effect that would be washed out in a
strong-coupling theory remains to be seen.  Finally, feature D is associated
with the phonon-related discontinuities in the dispersion at $\omega_{ph}$.
\par
The appearence of two gap-like features in the tunneling spectrum is not 
necessarily a consequence of two competing order parameters, but of gap 
anisotropy.  Indeed, for a pure (generalized) s-wave superconductor, there will
be two peaks in the tunneling dos whenever the gap is anisotropic, as long as
the minimum gap is non-zero.  Moreover, even for a pure d-wave superconductor, 
there can be two gap-like features if the VHS is not at the Fermi level,
Fig.\ref{fig:11}.

\subsection{Line Broadening}

In the one-dimensional CDW, the principal source of line broadening above the
Peierls transition is CDW fluctuations\cite{LRA,Sad,Tch}.  The phonon
propagator diverges at the transition:
\begin{equation}
D(q)\simeq{1\over\xi^{-2}+|\vec q-\vec Q|^2},
\label{eq:41}
\end{equation}
with $\xi^{-2}\propto (T-T_p)$.  For the two-dimensional Van Hove problem, a
similar form holds\cite{RM5} with $T_p=0$ in the absence of interlayer coupling.
This results in an intrinsic broadening $\Gamma =v_F/\xi\propto\sqrt{T}$.
\par
In the superconductor, a true long-range ordered state is possible, with
a real gap at the Fermi surface leading to a considerably reduced $\Gamma$.
Such a reduced scattering has been observed in a number of transport properties,
as well as in the photoemission spectrum.  However, the scattering should be
restored at high frequencies, when the CDW fluctuations can break a pair.  On
analogy with the results of Coffey and Coffey\cite{CoCo}, we assume that the
crossover will fall near $2\Delta_k$.
(Strictly speaking, the crossover should be at 3$\Delta_k$ for an s-wave
superconductor.)

\section{Comparison to Experiment}

\subsection{Experimental Situation}

Both photoemission and tunneling measurements of the pseudogap have been largely
restricted to Bi-2212.  While the general features of the spectra are reasonably
well understood, there are still differences in detail.  In this subsection, we
summarize a number of unresolved issues.

(a) In optimally-doped and lightly underdoped Bi-2212, photoemission detects a
significant rearrangement of spectral weight in the superconducting state.  The
low temperature ($T<T_c$) spectra are generally characterized by three features:
a sharp quasiparticle {\it peak} $\Delta_1$ separated from the Fermi level by a 
small ($\sim$ 25-50 meV) gap and a broad {\it hump} at much higher energies, 
$\Delta_2$ separated by a well-defined {\it dip} at energy $\sim 2\Delta_1$.  
Above $T_c$, the peak and dip disappear, leaving a feature similar to the hump.
\par
The dip feature is now widely interpreted as coupling to some form of collective
mode, the exact nature of which is not well understood.  The fact that it varies
with doping, scaling as $2\Delta_1$ is puzzling.  One possible interpretation is
that it is related to electron-electron scattering.  The broadening of the
photoemission features is known to decrease dramatically just below $T_c$, and
this decrease would be expected to fall off at frequencies much larger than the
gap (at $3\Delta$ for an s-wave superconductor, $2\Delta$ for d-wave).  Such a
result could be very useful in separating the superconducting gap from the 
pseudogap, but fits to this model have been unsuccessful in reproducing the dip 
amplitude.  We will show (Section III.E) that it is possible to get a large dip
if there is significant spectral weight in the hump feature.
\par
(b) The presence of the dip has led to some confusion on the doping dependence 
of `the gap': should one count $\Delta_1$ or $\Delta_2$.  The latter choice was
made by Marshall, et al.\cite{Gp0}.  In their earlier work, the Argonne group 
used neither choice, but defined the gap in terms of the leading edge near 
$\epsilon_F$.  This feature should scale with the peak position of $\Delta_1$,
but be somewhat smaller.  In their more recent work they have used the peak
position $\Delta_1$ for better comparison with tunneling studies.  Our analysis
suggests that this is a more appropriate choice, and we shall follow this latter
usage.  Our own belief is that, since the hump persists above $T_c$, $\Delta_2$
is a proper candidate for the pseudogap.  However, it is not always easy to
extract this feature from the published literature.  Fortunately, 
Ding\cite{DAPS} is now carrying out a careful study of both gaps.  
\par
Near optimal doping, $\Delta_1\simeq\Delta_2$, so the literature uncertainty 
will not greatly affect our reconstructed phase diagram, Fig.~\ref{fig:5},
below.  For stronger underdoping, the sharp peak and dip gradually wash out,
while the hump shifts to higher energy.  We find that this shift is clearly
revealed in the doping dependence of $T^*$, the temperature onset of the 
pseudogap\cite{DCN}, and that these data are consistent with the Stanford 
measurements of the hump data, suggesting that the photoemission provides a 
single pseudogap phase diagram.  Moreover, this diagram is remarkably consistent
with the pseudogap phase diagrams found for LSCO and YBCO on the basis of
transport measurements.
\par
(c) Tunneling is mainly sensitive to the sharp peak and dip feature in the
superconducting state, and it is not always clear whether the hump feature is
seen.  However, some groups clearly do see structure above $T_c$\cite{tu1}.
One problem is that tunneling is very surface sensitive, and the BiO layer on 
the top surface has an insulating gap.  However, consistency with the 
photoemission would suggest that a pseudogap should be present above $T_c$.  
Recently, Renner, et al.\cite{Ren} have reported that the tunneling gap inside 
a vortex core resembles that found in the pseudogap above $T_c$.
\par
(d) There is a question about overdoping, with Renner, et al.\cite{tu1} 
detecting the pseudogap in overdoped samples, while photoemission 
studies\cite{PEover} find a rapid collapse of the pseudogap in overdoped 
Bi-2212.  It seems likely, however, that in LSCO the pseudogap 
persists well into the overdoped regime\cite{Gp6}.
\par
(e) There is also some disagreement on the gap ratio $2\Delta (0)/k_BT^*$, with 
Oda, et al.\cite{tu4} reporting a value 4-5, whereas a value $\sim 8$ can be 
extracted from the data of Ding, et al.\cite{DCN}.

\subsection{Pseudogap Phase Diagram}

\par
\begin{figure}
\leavevmode
   \epsfxsize=0.33\textwidth\epsfbox{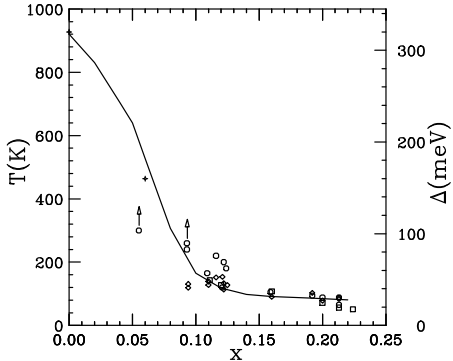}
\vskip0.5cm 
\caption{Pseudogap phase diagram in Bi-2212 determined from photoemission and
tunneling. Diamonds = photoemission gap\protect\cite{DCN}; +'s = photoemission 
gap\protect\cite{Gp0,Well}; circles = $T^*$ measured from 
photoemission\protect\cite{DCN}; squares = tunneling gap\protect\cite{tu3}.
Solid line = guide to the eye.}
\label{fig:5}
\end{figure}
\begin{figure}
\leavevmode
   \epsfxsize=0.33\textwidth\epsfbox{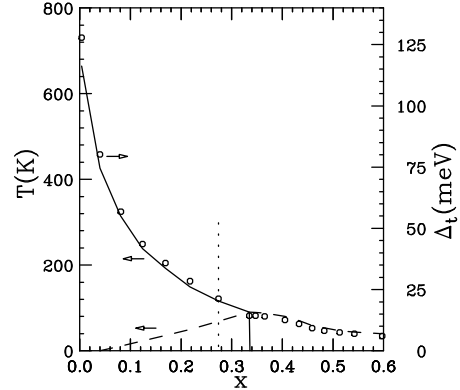}
\vskip0.5cm 
\caption{Model pseudogap phase diagram for Bi-2212. Solid line = CDW transition
$T_p$; dashed line = superconducting transition $T_c$; circles = total gap 
$\Delta_t$ at 1K.}
\label{fig:15}
\end{figure}

Figure~\ref{fig:5} summarizes the photoemission and tunneling data on the
pseudogap in Bi-2212.  Most of the photoemission data are taken from the 
Argonne group\cite{DCN}, but the two lowest dopings are from the Stanford 
group\cite{Gp0}, and from the insulating phase, Sr$_2$CuO$_2$Cl$_2$\cite{Well}. 
Since there has been disagreement on how to determine the pseudogap peak 
position, it is important to note that the Argonne data are consistent with a
large increase in $T^*$ (open circles) at low doping.  
Figure~\ref{fig:15} illustrates an optimized fit of the pBF model to the 
Bi-2212 pseudogap phase diagram.  Details about choosing the parameters are 
discussed in Appendix A.  Parameters are $t_0=0.25eV$, $x_c-0.335$, $\omega_
{ph}=35meV$, $\lambda_G=446meV$, $\lambda_{\Delta}=194meV$.
\par
\begin{figure}
\leavevmode
   \epsfxsize=0.33\textwidth\epsfbox{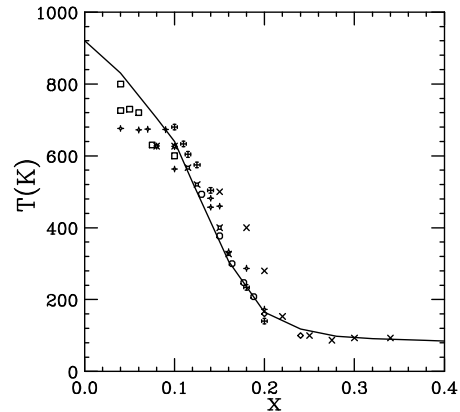}
\vskip0.5cm 
\caption{Pseudogap phase diagram in LSCO and YBCO determined from transport.
Open circles: from resistivity of YBCO; other symbols: transport measurements in
LSCO, see Ref.\protect\cite{BatT}; solid line from Fig.\protect\ref{fig:5}, with
$x$ axis scaled by a factor of 2.}
\label{fig:6}
\end{figure}

The overall qualitative and quantitative agreement is quite good.  A number of 
features of the experimental data are worth commenting on.  First, there is an 
approximately constant 
ratio between the total pseudogap, defined as the energy shift between the 
dispersion at $(\pi ,0)$ and the Fermi level, and the pseudogap onset 
temperature $T^*$, $2\Delta (0)/k_BT^*\simeq 8$.  A similar but smaller ratio,
4.1, is found in the calculation (note that this ratio is close to that found by
Oda, et al.\cite{tu4}).  Secondly, the overall shape of the curve,
strongly $x$-dependent in the underdoped regime, nearly constant in the 
overdoped regime, is well reproduced by the theory.  Thirdly, the pseudogap
determined from {\it photoemission} in Bi-2212 is in excellent agreement with
the pseudogap determined from {\it transport} in LSCO and YBCO\cite{BatT}, 
Fig.~\ref{fig:6}.  In all three cases, the doping axis had to be scaled to 
cause the curves to coincide.  The scaling suggests that, if the optimal
doping for superconductivity in LSCO is $x=0.16$, it is an effective $x=0.2$ in
YBCO, and $x=0.32$ in Bi-2212.  Such a shift in $x_c$ is consistent with the
Uemura plot\cite{Uem}.
\par
Finally, as noted previously\cite{tu3}, there is excellent agreement between
the photoemission pseudogap at $(\pi ,0)$, $\Delta_{PE}$, and the tunneling 
pseudogap, $\Delta_{TU}$, defined as half the splitting between the two 
tunneling peaks.  As illustrated in Fig.~\ref{fig:3}, this result is also in 
excellent agreement with theory.  Note that the theory holds specifically {\it 
when the Fermi level is pinned at the VHC}.  Indeed, we believe that the present
data\cite{tu3} constitute the strongest proof for this pinning.  To illustrate 
the strength of the evidence, in the following subsection we will test the null
hypothesis: assume that VH pinning is absent, and see how different $\Delta_{PE}
$ and $\Delta_{TU}$ would be.

\subsection{Van Hove Pinning vs d-wave Superconductivity}

\begin{figure}
\leavevmode
   \epsfxsize=0.33\textwidth\epsfbox{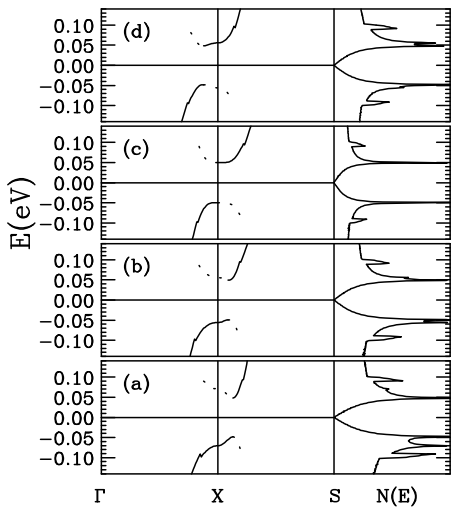}
\vskip0.5cm 
\caption{Energy dispersion near the Fermi level (left), and associated tunneling
density of states $N(E)$ (right) for a pure d-wave superconductor, as the VHS
sweeps through the Fermi level: $(E_{VHS}-E_F)/t$ = -0.2 (a), -0.1 (b), 0 (c), 
+0.1 (d).}
\label{fig:11}
\end{figure}

It has been suggested that what we have interpreted as Van Hove pinning can be 
alternatively explained as due to simple d-wave superconductivity, since the 
d-wave gap, $\Delta_d=\Delta_0(c_x-c_y)$ is largest near $(\pi ,0)$.  Here, we
demonstrate that this is not the case.  We assume that doping is accomplished
by the filling of a rigid band, Eq.~\ref{eq:1b}, and there is a fixed d-wave
gap with doping-independent strength $2\Delta_0=50meV$.  Figure~\ref{fig:11}
illustrates the energy band dispersion near the Fermi level, and the 
corresponding tunneling spectra, at a series of dopings away from the VHS.
The solid lines are the principal branches of the dispersion, with coherence 
factors $\ge 0.5$; the dashed lines are those parts of the ghost branches with
coherence factors in the range 0.1-0.5.  A large phonon energy $\hbar\omega_{ph}
=90$meV was assumed, to shift the phonon peaks in the tunneling spectra out of 
the range of interest.

The tunneling spectra clearly show the VHS moving through the Fermi
level.  Note that the VHS is always at $(\pi ,0)$, but the superconducting gap
is at the Fermi level, and hence in general away from $(\pi ,0)$.  
Superconductivity and the VHS actually produce two separate peaks in the
tunneling spectra.  (This fact has been noted earlier\cite{WeOst}.)  If the 
spectra are too broadened to resolve the individual 
peaks, one would find the lower (upper) peak to be more intense in underdoped 
(overdoped) samples.  This behavior is seen experimentally\cite{tu1,tu3}, and 
constitutes strong evidence that at optimal doping the Fermi level exactly
coincides with the VHS -- in agreement with an earlier prediction\cite{MG}.
We claim additionally that the Fermi level shifts away from the
VHS anomalously slowly on the underdoped side.

It can be seen that the dominant gap arises away from the VHS -- at the point
where the bands cross the Fermi level.  However, there is a subsidiary structure
associated with the saddle point dos peak, which rapidly shifts away from the
Fermi level with doping.  This would lead to a {\it double peak structure} in
the tunneling spectrum.  Moreover, it would mean that {\it the tunneling gap} 
(taken as the lower-energy, more prominent feature) {\it is distinct from the
photoemission gap} (defined by the dispersion at $(\pi ,0)$).  
\begin{figure}
\leavevmode
   \epsfxsize=0.28\textwidth\epsfbox{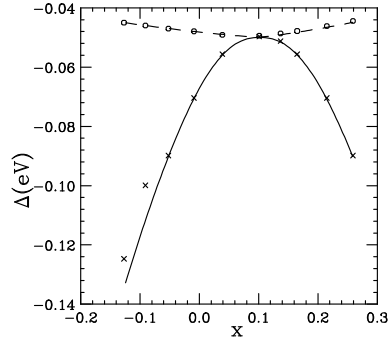}
\vskip0.5cm 
\caption{Doping dependence of $\Delta_{TU}$ the tunneling (circles and dashed 
lines) and $\Delta_{PE}$, the $(\pi ,0)$ photoemission ($\times$'s and solid 
curve) gaps, for a d-wave superconductor in the absence of VH pinning ($\tau=
-0.25$).}
\label{fig:12}
\end{figure}
\par
This is illustrated in Fig.~\ref{fig:12}, which shows the doping dependence of
these two gaps.  The symbols were estimated from the curves of Fig.~\ref{fig:11}
while the curves are analytical expressions, given by the curvature of the
dispersion relations (Appendix B).
\begin{figure}
\leavevmode
   \epsfxsize=0.28\textwidth\epsfbox{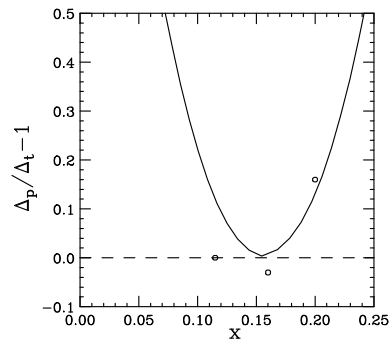}
\vskip0.5cm 
\caption{Normalized splitting of Fermi level from the VHS, measured as the 
normalized difference between the tunneling and photoemission pseudogaps, 
$(\Delta_{PE}-\Delta_{TU})/\Delta_{TU}$.  Solid line = theory in the
absence of pinning (similar to Fig.~\protect\ref{fig:12}, but with $\tau 
=-0.38$); open circles = derived from data of Refs.~\protect\cite{tu3,DCN}.}
\label{fig:13}
\end{figure}

\begin{figure}
\leavevmode
   \epsfxsize=0.28\textwidth\epsfbox{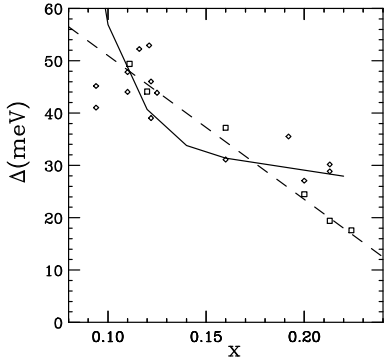}
\vskip0.5cm 
\caption{Comparison of photoemission (diamonds)\protect\cite{DCN} and
tunneling (squares)\protect\cite{tu3} gaps in Bi-2212.  Solid and dashed lines 
= guides to the eye.}
\label{fig:13a}
\end{figure}

In Figure~\ref{fig:13}, we plot the calculated difference between the tunneling
and photoemission gaps, normalized to the tunneling gap, and compare this to our
estimate of the {\it experimental} difference\cite{tu3}, Fig.~\ref{fig:5}.  
For convenience, we replot the tunneling and photoemission gaps from 
Fig.~\ref{fig:5} in Fig.~\ref{fig:13a}.
Only three data points could be extracted, and there are considerable error 
bars in the measurements.  (Note in particular that $\Delta_{PE}-\Delta_{TU}$ 
cannot be negative.)  The errors are smallest for the underdoped system:
both tunneling and photoemission data are reported in Ref.~\cite{tu3}, and are
presumably for similar samples.  In the other two systems, tunneling data from
Ref.~\cite{tu3} are compared with photoemission data from Ref.~\cite{DCN}.

Despite these limitations, the results are
intriguing.  There is a hint that the VHS depins from the Fermi level in the
overdoped regime, but there is no sign of depinning in the underdoped samples.
{\it Thus, the data appear to rule out the hypothesis of rigid band filling}, 
strongly implying that below optimal doping the Fermi level is pinned to the 
VHS.  Clearly, this experiment needs to be repeated much more carefully, and 
over a wider doping range.  In
particular, (a) the photoemission gap must be measured exactly at $(\pi ,0)$,
and (b) both photoemission and tunneling must be carried out on the same (or
identical) samples, to minimize sample-to-sample variations.  However, a
direct determination that the Fermi level is pinned to the VHS would confirm
a number of strong correlation theories, and would have a profound influence on
future theoretical modeling.  

\subsection{Tunneling Spectra}

\begin{figure}
\leavevmode
   \epsfxsize=0.33\textwidth\epsfbox{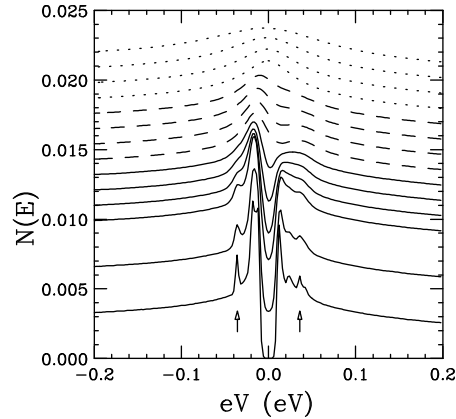}
\vskip0.5cm 
\caption{Tunneling dos for Bi-2212, using parameters of 
Fig.\protect\ref{fig:15} (dotted line).  From bottom to top, the temperatures
vary from 10K to 100K, in 10K intervals, then go to 120, 150, 200, and 300K.
Dotted lines, $T>T_p=116K$; dashed lines, $T_p>T>T_c=69K$; solid lines, $T<T_c$.
All curves are offset for clarity (all essentially coincide for $|eV|\ge 
100meV$).}
\label{fig:16}
\end{figure}

Figure~\ref{fig:16} illustrates the evolution of the calculated tunneling 
spectra with temperature in the underdoped regime ($x=0.274$, dotted line in 
Fig.~\ref{fig:15}).  The dotted lines show the normal state VHS above $T_p$;
the dashed lines are in the CDW phase; and the solid lines are in the mixed
CDW-superconducting phase.  The only broadening is thermal, due to the Fermi
function; the disorder broadening $\Gamma=0$.  As noted above 
(Fig.~\ref{fig:3}b), there are prominent features at the phonon frequency,
$\omega_{ph}=35meV$ (arrows).  For a realistic phonon spectrum, they would split
up into the strong-coupling factor $\alpha^2F$.  There have been a number of 
reports in the literature of the observation of such structure, but it is not
clear how reproducible it is.
\begin{figure}
\leavevmode
   \epsfxsize=0.33\textwidth\epsfbox{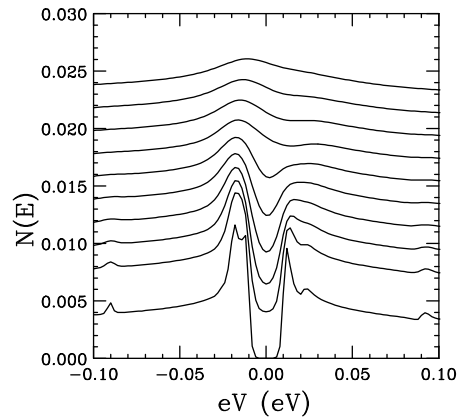}
\vskip0.5cm 
\caption{Tunneling dos for Bi-2212, using parameters of 
Fig.\protect\ref{fig:16}, but with $\hbar\omega_{ph}$=90meV.  From bottom to 
top, the temperatures vary from 10K to 100K, in 10K intervals.}
\label{fig:26}
\end{figure}
\par
\begin{figure}
\leavevmode
   \epsfxsize=0.33\textwidth\epsfbox{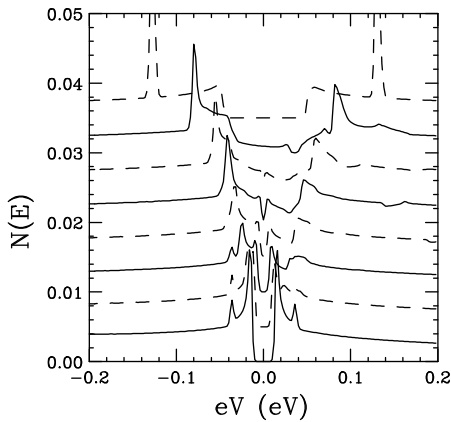}
\vskip0.5cm 
\caption{Tunneling dos for underdoped Bi-2212 at 10K, as a function of
doping, using parameters of Fig.\protect\ref{fig:15}.  From top to
bottom, the doping is $x$ = 0.004, 0.04, 0.08, 0.124, 0.169, 0.218, 0.274, and
$x_c=0.34$.  All curves are offset for clarity (all approximately coincide at 
$eV=200meV$).}
\label{fig:17}
\end{figure}

In order to better see the shape of the tunneling peaks, the dos is recalculated
in Fig.~\ref{fig:26}, with all parameters unchanged except $\hbar\omega_{ph}
=90$meV.  At 10K and 20K, a splitting of the principal dos peaks is just 
resolvable.  This 
splitting is due to the presence of combined CDW and superconducting order, but
it does {\it not} simply mean that the two phases have separate spectral peaks.
Thus, at $(\pi ,0)$, there is a single peak, at the position given by 
Eq.~\ref{eq:1c}.  Instead, the two gaps have different dispersion, with the
superconducting gap always at $\epsilon_F$ and the CDW gap dispersing away from
$\epsilon_F$ (see feature C in Fig.~\ref{fig:3}b).  Thus, at $(\pi ,0)$ the
gap is the (Euclidean) sum of the CDW and superconducting gap, while near
$(\pi /2,\pi /2)$ it is only the latter.  The splitting of the peaks in
Fig.~\ref{fig:16} reflects this anisotropy.  This splitting is 
essentially lost by 30K, due to thermal broadening, and would not survive much
disorder broadening.  Note that the gap fills in with increasing temperature,
without significantly closing.  Indeed, since superconductivity disappears at
a lower temperature, the gap actually appears to increase at higher
temperatures.  The low-temperature splitting increases with increased 
underdoping (Fig.~\ref{fig:17}), and hence might ultimately be observable 
experimentally.  However, the data of Fig.~\ref{fig:16} correspond to $T_c=69K$,
somewhat lower than any reported tunneling data.  Moreover, the calculations are
for an s-wave superconductor, and the smaller gap could be very different, or
even absent, for a d-wave superconductor.
\par
Thus, the model agrees with experiment\cite{tu4} in finding gap-like features 
above $T_c$, but does not reproduce the rearrangement of features below $T_c$
into a sharp peak, dip, plus hump.  A suggested explanation for these features
is given in the following Section.
\par
Figure~\ref{fig:17} shows the evolution of the low-temperature tunneling spectra
with doping, in the underdoped regime.  At optimal doping (lowest curve) the gap
is symmetrical and pure superconducting.  As the material is successively
underdoped, the peaks split, with the pseudogap peak growing and the 
superconducting peak shrinking, scaling with $T_c$, down to $x\simeq 0.06$ where
superconductivity disappears.
\par
\begin{figure}
\leavevmode
   \epsfxsize=0.28\textwidth\epsfbox{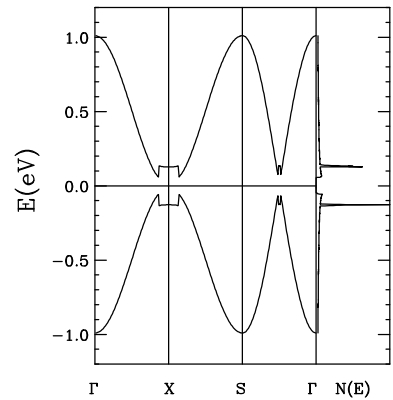}
\vskip0.5cm 
\caption{Band dispersion near half filling Bi-2212 at 10K, using parameters of 
Fig.\protect\ref{fig:15}.}
\label{fig:18}
\end{figure}

The lowest doped state, close to half filling, is very interesting.  For this 
state the present model is expected to be least accurate.  The state is
dominated by an electronic instability, the Mott transition.  There is a 
charge-transfer gap of $\sim 1.6eV$, and a residual dispersion suggestive of the
flux phase.  Figure~\ref{fig:18} shows that both these features are 
qualitatively  present in
the model.  From Eq.~\ref{eq:11b}, it can be seen that near half filling, when
$t^{\prime}\rightarrow 0$, the CDW gap opens up over the full Fermi surface,
while the excess $G_1$ gap ensures that the largest gap is present near the VHS.
The chief differences from experiment are (a) the gap evolves discontinuously
from $G_0$ to $G_0+G_1$ rather than smoothly, (b) the overall bandwidth is 
too large, since the model has not taken into account that correlations 
renormalize the bandwidth from $\sim 8 t$ to $\sim 2J$\cite{Pstr}, and (c)
the model cannot reproduce the magnitude of the (electronic) Mott gap, $\sim
1.6eV$.  Indeed, the full gap is limited to $\le 2\hbar\omega_{ph}$.  

\subsection{Peak, Dip, and Hump}

\begin{figure}
\leavevmode
   \epsfxsize=0.33\textwidth\epsfbox{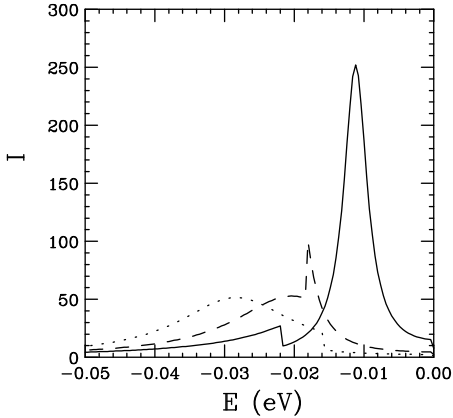}
\vskip0.5cm 
\caption{$(\pi ,0)$ photoemission spectra for Bi-2212 at T=1K, using parameters 
of Fig.\protect\ref{fig:15} (x=0.22), except $\lambda_G$ = 0.446 (dotted line),
0.385 (dashed line), and 0.325eV (solid line).  Broadening given by 
Eq.~\protect\ref{eq:32}.} 
\label{fig:27}
\end{figure}

The photoemission in Bi-2212 near $(\pi ,0)$ has a characteristic shape in the
superconducting state which is difficult to reproduce in the current model.
In the pseudogap state above $T_c$, there is a very broad hump, consistent
with a CDW gap with large broadening $\Gamma$, as in Section II.D.  In the
superconducting state, the hump remains, but the low-energy side of the hump is 
transformed into a very sharp feature at energy $\Delta_1$, with a dip at energy
$\sim 2\Delta_1$.  In the CDW superconductor, there is not expected to be such 
a two-peaked structure along $(\pi ,0)$, but a single peak given by 
Eq.~\ref{eq:1c}.  However, as noted in Section II.D, the broadening can have a
strong energy dependence, with a long lifetime in the superconducting state
cut off near $2\Delta$\cite{CoCo}.  In the absence of a full theory, it is not
clear which gap $\Delta$ represents, but since this is specifically a coherence
effect, it seems reasonable to assume that it should be approximately the
superconducting $\Delta_k$.  Since $\Delta_k<\Delta_t$, one can get a behavior
at low temperatures which is closer to experiment.  Figure~\ref{fig:27} 
illustrates the three possible behaviors, depending on whether $2\Delta_k$ is
$<<$, $\sim <$, or $>$ $\Delta_t$: one can get hump plus dip (dotted line), hump
plus peak (dashed line), or peak plus dip (solid line).  In the last case, the
dip is not necessarily at twice the peak in energy.
\begin{figure}
\leavevmode
   \epsfxsize=0.33\textwidth\epsfbox{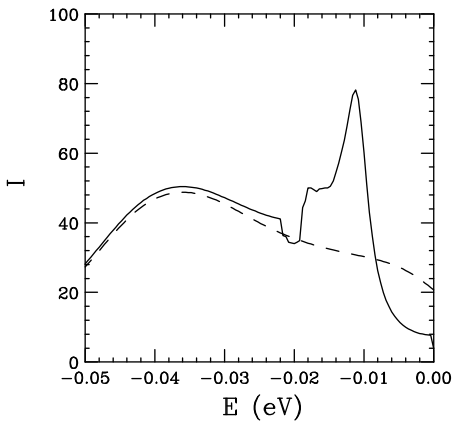}
\vskip0.5cm 
\caption{$(\pi ,0)$ photoemission spectra for Bi-2212, using parameters 
of Fig.\protect\ref{fig:15} (x=0.22), except inhomogeneously broadened (averaged
over 20 $\lambda_G$ values).  Solid line: T = 1K, dashed line: T = 70K.}
\label{fig:28}
\end{figure}
\par
It seems clear, however, that the hump is associated with the pseudogap, and the
peak with the superconducting gap.  We suggest that the experimental result may
represent an effect of phase separation, which is not well captured by our 
model.  To test this hypothesis, we developed a simple model, assuming that
the pseudogap feature is inhomogeneously broadened.  Specifically, we assume
that $\lambda_G$ is spatially inhomogeneous, and that regions of different
$\lambda_G$ evolve independently.  For each region, lowering the temperature
will produce a spectrum like one of those in Fig.~\ref{fig:27}.  In the regions
of large $\lambda_G$, superconductivity has little effect, leaving the hump
unchanged.  However, when $\lambda_G$ is small, at low T there will be a sharp
superconducting peak plus well-defined dip.  Figure~\ref{fig:28} shows the
resulting spectra, both just above the onset of superconductivity, at T=70K
(dashed line), and at low temperature, T=1K (solid line).  Each spectrum is the
superposition of 20 individual spectra, each with
\begin{equation}
\Gamma=\cases{\Gamma_1,&if $|E|\le 2\Delta_k$
                      \cr
                       \Gamma_2,&otherwise,\cr}
\label{eq:32}
\end{equation}
with $\Gamma_1=2meV$, $\Gamma_2=10meV$.  The general features of the experiment 
are clearly reproduced, although the sharp peak is not resolution limited, due
to the inhomogeneous broadening.

This interpretation of the sharp peak as a measure of $\Delta_k$ leads to a new
problem, since the experiments imply that this $\Delta_k$ increases with 
decreasing $x$.  To reproduce this behavior in the phase diagram would require
that $\lambda_{\Delta}$ varies with doping -- the same effect needed to explain
the LSCO phase diagram, Fig.~\ref{fig:21}.  Such a doping dependence would be
consistent with magnetic fluctuation-induced superconductivity.  However, it
must be kept in mind that an alternative interpretation, such as the solid line
in Fig. \ref{fig:27}, was ruled out because the dip feature falls at exactly
$2\Delta_1$.  There exist some data (e.g., Fig. 1 of Ref. \cite{Wei}) in
which the dip falls at an energy substantially less than $2\Delta_1$.  This
discrepancy must be clarified before a definitive model for the sharp peak can
be established.

\subsection{The Incredible Shrinking Fermi Surface}

Norman, et al.\cite{NoD} have shown that the Fermi surface in the pseudogap 
phase has a remarkable temperature dependence: there is a full, large Fermi
surface above the pseudogap transition, $T_p$, but as $T$ is reduced below 
$T_p$, the Fermi surface gradually collapses -- vanishing first near $(\pi ,0)$
and then over a larger angular range before ultimately being reduced to a point
along the line $c_x=c_y$.  This low temperature limit is consistent with a 
d-wave superconductor, but the intermediate temperature regime is not: the gap
should open at $T_c$ everywhere on the Fermi surface, with the magnitude of the
gap proportional to $(c_x-c_y)$.  Fluctuation broadening could produce this
effect by limiting the pseudogap to those parts of the Fermi surface where it is
larger than the energy uncertainty\cite{NoRD,GIL,Naz2}.  
However, it can also be interpreted in terms of a VH nesting gap.
\par
\begin{figure}
\leavevmode
   \epsfxsize=0.33\textwidth\epsfbox{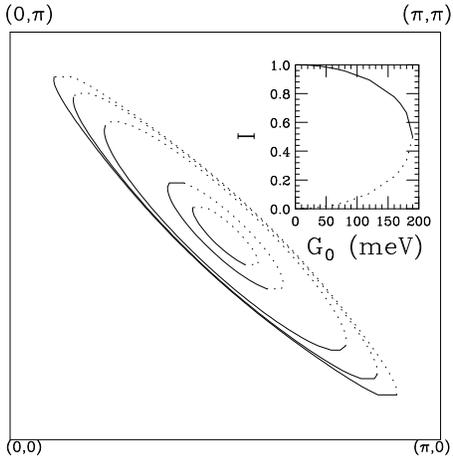}
\vskip0.5cm 
\caption{Fermi surface in CDW phase, for $\tau =-0.38$, $G_0$ = 20, 40, 80, 
160, and 180 meV, for progressively smaller ellipses.  Dotted segments = ghost
Fermi surfaces, with coherence factors $< 0.5$. Inset: coherence factors along
the diagonal ($k_x=k_y$).}
\label{fig:7}
\end{figure}
\begin{figure}
\leavevmode
   \epsfxsize=0.33\textwidth\epsfbox{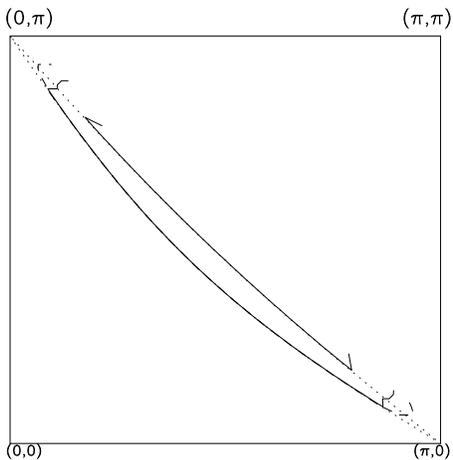}
\vskip0.5cm 
\caption{Fermi surfaces in CDW phase, for self-consistent parameters.  Lower
(upper) set of lines: $\tau$ = -0.4 (-0.16), $\lambda_G$ = 0.50 (0.25) eV.
Dotted lines = in normal state above $T_p$; solid lines = in pseudogap state, 
just above $T_c$; dashed lines = intermediate temperatures.  Only segments of
Fermi surface with coherence factors $\ge 0.1$ are shown.}
\label{fig:7a}
\end{figure}

Figure~\ref{fig:7} shows the evolution of the Fermi surface with $G_0$, for
$\Delta =G_1=0$.  In a pure CDW state, the Fermi surface can be 
calculated analytically, Appendix C.  For finite $G_0$, the surface is an 
ellipse, with
the inner half derived from the original Fermi surface and the outer half a 
ghost Fermi surface, zone folded by $Q$.  The latter has considerably lower 
intensity (inset), and will be further reduced
by fluctuation effects and stripe effects; it is not seen experimentally.  On
the other hand, the truncated original surface has a large coherence factor 
($\ge 0.5$), and bears a strong resemblance to the experimental data.  
Below $T_c$, the $G_i$'s shrink
as $\Delta$ grows, so the opening of the superconducting gap takes over.  Thus,
for a d-wave superconductor, the residual points of zero gap would lie along
the original Fermi surface, as observed, and not at $(\pi /2,\pi /2)$.
When
$G_1\ne 0$, Fig.~\ref{fig:7a}, the resemblance to experiment is even closer:
the Fermi surface shrinks down with essentially no change of shape, from the
pseudogap onset to the superconducting $T_c$. 
\par
There is one possibly significant difference: as the normal state above $T_p$
is approached, Fermi surfaces should approach the VHS's at $(\pi ,0)$, which is
not clearly seen in experiment.  A related problem is the minimum gap locus
found in the pseudogap phase.  These problems will be discussed further in the 
following section.

\subsection{Minimum Gap Locus}

\begin{figure}
\leavevmode
   \epsfxsize=0.33\textwidth\epsfbox{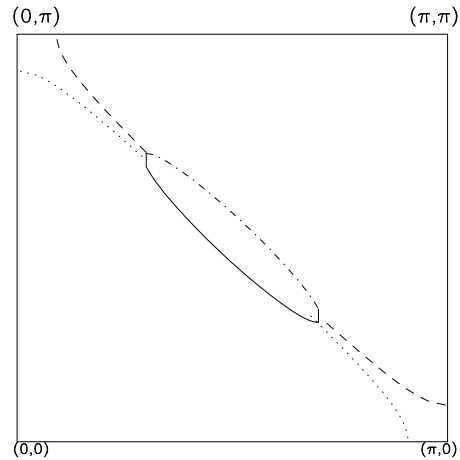}
\vskip0.5cm 
\caption{Loci of minimum gap (dashed and dotted lines) and zero gap 
(solid and dot-dashed lines).}
\label{fig:8}
\end{figure}

In order to recover a `Fermi surface' in the presence of a (pseudo)gap, Ding,
et al.\cite{DgNY} have introduced the concept of a `locus of minimum gap'.
For a series of cuts in the Brillouin zone, they define the minimum gap -- the
$\vec k$-point at which the photoemission gap most closely approaches the Fermi
level.  The gap is defined as the leading edge of the photoemission pattern,
but the comparison with tunneling\cite{tu3} suggests that the peak position 
should work as well.  The Fermi surface would then correspond to a locus of
zero-gap excitations.  In Appendix C, analytical expressions for
the Fermi surface and minimum gap locus (actually, the maximum in the dispersion
of the lower band) are derived, which are plotted as lines in Fig.~\ref{fig:8}. 
The solid lines are the Fermi surface, the dashed lines the minimum gap loci, 
and the dot-dashed and dotted lines their ghostly counterparts.  Neglecting the 
ghost surfaces, the resulting dispersion resembles the 
experimental findings\cite{DgNY}.  Note in particular that the locus of minimum
gap does {\it not} pass through the $(\pi ,0)$ point, even though $\epsilon_F$ 
is pinned at the VHC (see Eq.~\ref{eq:B4}).  This can readily be understood: 
since the gap is largest at $(\pi ,0)$, the scans find a smaller gap at some 
$k_x$ away from this point.  Knowing the full dispersion $E(\vec k)$, a much 
fuller test of the theory can in principle be carried out.

While the problem of the minimum gap locus can be satisfactorily explained in
the pinned VHS model, the problem remains that in the normal state above the 
pseudogap, the Fermi surface found in photoemission does not seem to intersect
$(\pi ,0)$.  It must be recalled that the VHS is the ultimate hot spot, so
local residual disorder is likely to broaden or split the quasiparticle spectrum
near the VHS even at temperatures considerably in excess of the pseudogap
temperature.  An indication for such smearing is the disagreement between the 
Argonne\cite{DgNY} and Stanford\cite{Gp0} groups on the shape of the Fermi 
surface near the VHS in underdoped cuprates.  Hence we prefer to infer
the position of the VHC from data in the superconducting state, when line
broadening effects are smallest.  If the VHC is within 5meV of $\epsilon_F$ in
the superconducting state (Fig. \ref{fig:13}), it should be even closer in the 
normal state, since the superconducting transition tends to
shift the chemical potential away from the VHS\cite{MFN}.

\subsection{Spectral Weight Shift}

Shen, et al.\cite{ZX} have recently found significant shifts of spectral weight
on cooling the sample from the pseudogap phase above $T_c$ to the 
superconducting phase below.  These are exactly the sort of changes one would
expect in the present model, since the CDW and superconductivity have very
different gap structures, and the CDW gap is significantly {\it reduced} below
$T_c$, Fig. \ref{fig:1d}.  However, experimentally these changes are found to
occur over an energy range extending to $\sim 300 meV$ away from $\epsilon_F$.
This extended range is presumably telling us about the relevant pairing bosons,
but since these energies are greater than $\hbar\omega_{ph}$, strong coupling
calculations are required, which are beyond the scope of the present paper.

\section{Conclusions}

We have taken a simple pinned Van Hove {\it Ansatz} for the striped pseudogap
phase in the cuprates, and analyzed the predicted tunneling and photoemission
spectra, comparing them with experimental data.  Highlites of our results
include: (1) We explain the experimental observation that the tunneling peaks 
coincide with the $(\pi ,0)$ photoemission dispersion, and show that careful
measurements of this effect can provide direct evidence of pinning of the
Fermi level to the VHC over an extended doping range.  (2)  In turn, the fact 
that tunneling shows a well-defined gap-like feature confirms that the $(\pi
,0)$ dispersion also has two branches -- that is, that the pseudogap is
associated with some form of VHS nesting\cite{Pstr}.  (3) The tunneling gap has
a characteristic asymmetry which vanishes at optimal doping; this is evidence
that optimal doping is that point at which the Fermi level exactly coincides
with the VHS\cite{MG}.
\par
(4) By plotting the doping dependence of the photoemission hump feature, the
resulting pseudogap phase diagram for Bi-2212 is in good agreement with similar
phase diagrams for LSCO and YBCO, derived from transport measurements.  (5)
The rearrangement of spectral weight seen below $T_c$ can be interpreted as
a generation of separate pseudogap (hump feature) and superconducting (sharp
feature) peaks with the dip between them due to superconducting coherence 
effects.  However, this requires inhomogeneity not included in the simple {\it
Ansatz}.  (6) Finally, a number of specific features observed in the 
photoemission receive a natural explanation in this model, including the
shrinking Fermi surface and the locus of minimum gap.
\par
A shortcoming of the model is the prediction of a splitting of the tunneling 
gap, which has not yet been observed.  However, we note that (a) this may be
due to the assumption of s-wave superconductivity, and (b) even the assumption 
of a pure d-wave superconductor, with no pseudogap, would lead to a split peak
in tunneling unless the Fermi level were pinned to the VHS.

\section{Discussion}

\subsection{Alternative Pseudogap Scenarios}

The present interpretation of the pseudogap in terms of Van Hove nesting is in
agreement with some early (1990) calculations\cite{KaSch,RM5} which predated
any of the experimental observations of the `spin gap' or pseudogap.  However,
one of these theories involves an SDW\cite{KaSch}, the other a CDW\cite{RM5},
while more recent calculations\cite{Laugh,WeL,Pstr} involve a flux phase near 
half filling.  Hence, the essential feature is the Van Hove nesting; the
particular instability depends on the details of the electron-electron
interaction.
\par
It is convenient to classify these instabilities in terms of the SO(6) 
scenario.  As discussed in Section II.B, the instabilities fall into two broad 
classes: nesting and pairing.
We expect that the phase diagram, Fig.~\ref{fig:1},
of the competition between pseudogap and superconductivity, would not
be greatly changed if the pseudogap represents any {\it nesting} instability,
{\it as long as the physics is dominated by proximity to a VHS}.  This would
hold for any Hubbard or tJ model, in which the VHS is at half filling, and
more particularly for any model which introduces a higher order hopping
parameter ($t^{\prime}$ or $t_{OO}$) to move the VHS toward finite positive $x$
(hole doping) -- i.e., closer to optimal doping.  
\par
Among the {\it nesting} theories of the pseudogap, we include Schrieffer's spin 
bag\cite{KaSch} and Zhang's SO(5)\cite{Zhang5}, for which the instability is
an SDW, Laughlin\cite{Laugh}'s and Wen and Lee\cite{WeL}'s flux phase models,
and Klemm's CDW model\cite{RAK}.  Within an SO(6) symmetric model, all would
produce the same pseudogap phase diagram.  Hence, the question of which one
actually produces the ground state depends on which symmetry-breaking operators
are present in the physical cuprates.  
\par
Following the discussion in Section II.B, it would be more difficult to
reproduce the pseuodgap phase diagram in terms of a {\it pairing instability} --
i.e., assuming that the pseudogap is a signature of superconducting 
fluctuations, a precursor of real space pairing.  We believe that this class of 
models will have difficulty explaining striped phases, and the seemingly smooth 
extrapolation of the pseudogap to the SCOC dispersion at half filling.  

\subsection{Distinguishing the Nesting Instabilities}

\par
There is a natural generalization\cite{KMV} of Eq.~\ref{eq:1c} to SO(6):
\begin{equation}
\Delta_t=\sqrt{\sum_{i=1}^{12}\Delta_i^2},
\label{eq:21c}
\end{equation}
where the sum is over all twelve instabilities of both superspins.  Thus, {\it
at $(\pi ,0)$ the pseudogap depends only on the vector sum of the 
individual gaps}.  For $t^{\prime}=0$, Eq. \ref{eq:21c} holds over the full 
Fermi surface.  Hence, the instabilities can only be distinguished by their
dispersion away from $(\pi ,0)$.

In attempting to distinguish different models, it must be kept in mind that this
secondary dispersion is likely to be model dependent.  For example, d-wave
superconductivity would have different dispersion depending whether it was
phonon-induced or an electronic instability.  In particular, the dispersion will
be sensitive to the nature of the boson mediating superconductivity, in 
particular to the cutoff, $\omega_c$.  
\par
A second example is instructive.  In the antiferromagnetic state at half filling
there is a competition between the N\'eel phase (commensurate SDW) and the flux
phase.  In a Hartree-Fock calculation, both phases give rise to similar gaps at
$(\pi,0)$, but very different dispersion in other directions.  In particular,
the N\'eel gap is approximately isotropic, while the flux phase has zero gap
near $(\pi /2,\pi /2)$.  (See Fig. 32 of Ref.~\cite{Surv} and the associated
discussion for references to earlier literature.)  Having the larger average 
gap, the N\'eel phase is more stable, but the experimental dispersion of 
SCOC\cite{Well}, and it was speculated that the N\'eel phase dispersion is
modified by fluctuation or correlation effects.  This seems to be the case,
since calculations of the dispersion of one hole in an antiferromagnet are
capable of reproducing the experimental dispersion\cite{Naz}.  Thus, while the
theoretical interpretation has proven quite involved, the dispersion remains 
simple: it has the preiodicity of the N\'eel superlattice, with maximal gap
associated with VHS splitting.
\par
Despite these caveats, we are 
exploring an SO(6)-generalized version of the pBF model, to test how sensitive 
the phase diagram and tunneling and photoemission spectra are to the particular
forms of nesting or pairing instabilities.

\subsection{Doping Dependence of $\lambda_{\Delta}$}

There were a few tantalizing hints that the superconducting coupling parameter
$\lambda_{\Delta}$ is doping dependent: increasing as $x$ is reduced below 
optimal doping.  This was needed to explain the LSCO phase diagram,
Fig.~\ref{fig:21}, and the doping dependence of the sharp photoemission peak
below $T_c$, Section III.E.  This could have a simple explanation -- e.g., the
bandwidth is reduced by correlation effects from $\sim 4t$ (half bandwidth) at
optimal doping to $\sim 2J$ near half filling, and this reduction is neglected 
in the pBF model.  Alternatively, it could signal the importance of magnetic
fluctuations.

In the three-band slave boson model calculation\cite{Pstr}, the stripes were
interpreted as a crossover from a flux phase near half filling to a CDW phase
near optimal doping.  In SO(6), this is a crossover between the two different
superspin multiplets, and is accompanied by a crossover from d-wave to s-wave
superconductivity.  This would be consistent with hints of a more s-wave-like
superconducting transition in overdoped materials.

We would like to thank NATO for enabling A.M. Gabovich to visit us and discuss
his work. Publication 744 of the Barnett Institute.

\appendix
\section{Phase Diagram for LSCO and YBCO}

We briefly discuss the procedure for determining the phase diagram of 
Figs.~\ref{fig:1}-\ref{fig:21} and \ref{fig:15}, to indicate the role of the 
various parameters.

{\bf 1. Choice of $\tau$.}  Using Eq.~\ref{eq:2}, a value $\tau =2t^{\prime}/t
=-0.38$ is necessary for the VHS to fall at optimal doping $x_c=0.16$ in LSCO.
On the other hand, there is a suggestion\cite{Mor} that $\tau$ in LSCO may be as
large as -0.6, which corresponds to $x_c=0.274$.

In YBCO, band structure calculations suggest $\tau =-0.9$ -- remarkably close to
the critical value -1.0 for one-dimensional behavior.  This in turn corresponds 
to a doping $x_c=0.54$.  Note however, that the band structure calculations also
find a significant interplane coupling between the two CuO$_2$ planes in a unit 
cell, such that the symmetrical combination is essentially undoped.  In this
case, for the antisymmetrical combination Fermi surface to be at the VHS would
require an average doping per layer of 0.54/2=0.27 holes.  This is consistent 
with the optimal doping estimated from the Uemura plot\cite{Uem}, which finds
that $x_c/m$ is about twice as large for YBCO as for LSCO.  Assuming that both
materials have the same effective mass $m$, this would give a broad peak in 
$x_c$ near 0.32 for YBCO.

The situation for Bi-2212 is slightly more complicated.  The Uemura plot
suggests an $x_c$ similar to that in YBCO, while photoemission is consistent
with $\tau\simeq -0.9$\cite{DgNY}.  However, there is no evidence in 
photoemission for the interlayer splitting of the Fermi surface.  We will
postulate that a splitting similar to that found in YBCO is locally present, but
is obscured by a very short c-axis correlation length.

{\bf 2. Optimal Doping.}  In YBCO, the pseudogap appears to vanish very close to
optimal doping, as soon as the superconducting $T_c$ exceeds the pseudogap 
$T_p$.  This is a natural consequence of the BF model, and so is a convenient
point to fix for the phase diagram.  We proceed as follows.  For a given $\tau$,
we set the Fermi level at the VHS and calculate the electron-phonon coupling 
energies, $\lambda_{\Delta}$ and $\lambda_G$ which give $\Delta_k=G_k=0^+$ at 
$T=90K$.  With this choice of $\lambda$'s, $T_c=T_p=90K$ at this doping, and at 
any larger doping the pseudogap will be suppressed to zero by $T_c$.  
\begin{figure}
\leavevmode
   \epsfxsize=0.3\textwidth\epsfbox{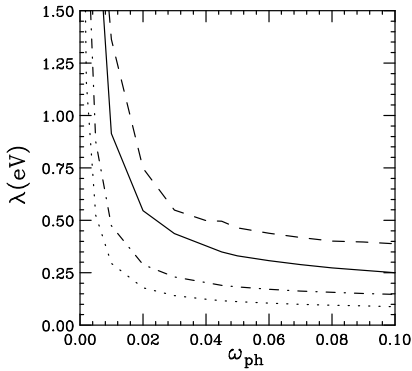}
\vskip0.5cm 
\caption{Allowed values of $\lambda$ vs $\omega_{ph}$ to fix $T_p=T_c=90K$ at
optimal doping.  Solid (dashed) line = $\lambda_G$ for $\tau$ = -0.6 (-0.9);
dotdashed (dotted) line = $\lambda_{\Delta}$ for $\tau$ = -0.6 (-0.9).}
\label{fig:22}
\end{figure}

The only free parameter is $\omega_{ph}$.  In Fig.~\ref{fig:22},
we plot the resulting $\lambda$'s vs $\omega_{ph}$ for two choices of $\tau$.
As expected, for a larger $\omega_{ph}$, a smaller $\lambda$ is needed, but for
$\omega_{ph}>30meV$, the value of $\lambda$ varies only weakly.  The dependence
of the $\lambda$'s on $\tau$ is what would be expected from Fig.~\ref{fig:1}:
nesting is strongest for $\tau =0$, so the necessary $\lambda_G$ increases with
increasing $\tau$, while pairing becomes easier as the bands become more
one-dimensional, $\tau\rightarrow 1$.
\begin{figure}
\leavevmode
   \epsfxsize=0.3\textwidth\epsfbox{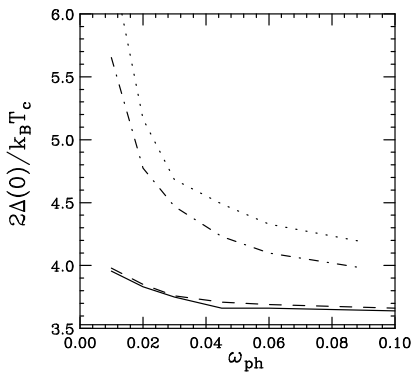}
\vskip0.5cm 
\caption{Gap ratio vs $\omega_{ph}$ using values of $\lambda$ determined from 
Fig.~\protect\ref{fig:22}.  Solid (dashed) line = ratios for superconducting gap
for $\tau$ = -0.6 (-0.9); dotdashed (dotted) line = ratios for CDW gap for 
$\tau$ = -0.6 (-0.9).}
\label{fig:23}
\end{figure}

{\bf 3. Choosing $\omega_{ph}$}  If we assume $t=0.25eV$ is known, and that
$\lambda_{\Delta}$ and $\lambda_G$ are doping independent, there is only one 
remaining free parameter, $\omega_{ph}$, to adjust to fit the rest of the
phase diagram.  We explore three features in particular: the gap ratio, the
magnitude of the limiting pseudogap at zero doping, and the superconducting 
onset.  

In weak coupling BCS
theory, the gap ratio $2\Delta (T=0)/k_BT_c=3.53$, and increases for stronger
coupling.  Figure~\ref{fig:23} shows the variation of this ratio with 
$\omega_{ph}$, where the $\lambda$'s are fixed by Fig.~\ref{fig:22}, as
discussed above.  The ratio for the superconducting phase is somewhat enhanced 
from the BCS value, and depends only weakly on $\omega_{ph}$.  The ratio can
be much larger and more $\omega_{ph}$-dependent for the pseudogap.  Note that in
Fig.~\ref{fig:1}, where $\lambda_{\Delta}=\lambda_G$ is assumed, the two gap
ratios are nearly equal.  (The problem with Fig.~\ref{fig:1} is that the
critical doping is too low.)  Experimentally, Oda, et al.\cite{tu4} find a value
4-5, whereas the data of Ding, et al.\cite{DCN} yield $\sim 8$. Thus, the model
can account for the former value, but not the latter.  If the correct ratio
turns out to be $\sim 8$, this may be evidence for pairbreaking effects limiting
$T_c$\cite{pair}.
\begin{figure}
\leavevmode
   \epsfxsize=0.28\textwidth\epsfbox{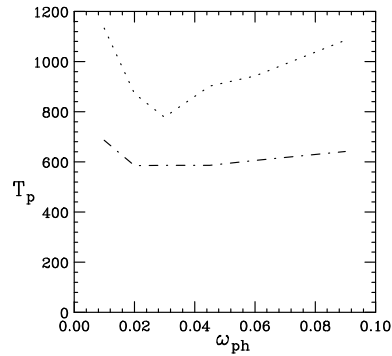}
\vskip0.5cm 
\caption{CDW onset $T_p$ at zero doping vs $\omega_{ph}$ using values of 
$\lambda$ determined from Fig.~\protect\ref{fig:22}.  Dotdashed (dotted) line 
= $T_p$ for CDW gap for $\tau$ = -0.6 (-0.9).}
\label{fig:24}
\end{figure}

Figure~\ref{fig:24} shows how the limiting pseudogap $T_p(\tau =0)$ depends on
$\omega_{ph}$ (in view of the uncertainty in the gap ratio, we have chosen to
work with the critical temperature rather than the gap).

\begin{figure}
\leavevmode
   \epsfxsize=0.28\textwidth\epsfbox{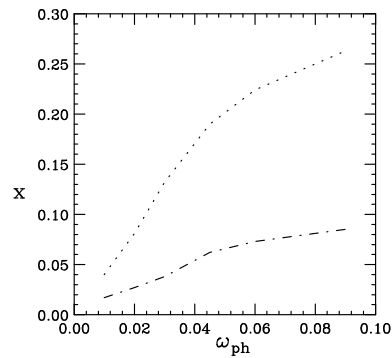}
\vskip0.5cm 
\caption{Onset of superconductivity $x$ vs $\omega_{ph}$ using values of 
$\lambda$ determined from Fig.~\protect\ref{fig:22}.  Dotdashed (dotted) line = 
$x$ for CDW gap for $\tau$ = -0.6 (-0.9).}
\label{fig:25}
\end{figure}

Another constraint is the superconducting onset $x_{min}$, the doping at which
superconductivity first appears.  This is $\sim 0.05$ in LSCO; for YBCO and
Bi-2212, it is hard to specify the precise planar hole doping.  Adjusting 
$x_{min}$ is somewhat in conflict with the other constraints, since a large
difference $\lambda_G-\lambda_{\Delta}$ leads to a larger $x_{min}$, 
Fig.~\ref{fig:25}.  

Given these conflicting constraints, the parameters of Fig.~\ref{fig:15} seem
to provide the best overall solution.

\section{Analysis of d-wave Gaps}

For a purely superconducting gap ($G_0=G_1=0$) the eigenvalues of Eq.~\ref{eq:3}
simplify:
\begin{equation}
E_{\pm}=\pm\sqrt{(\epsilon_k-\epsilon_F)^2+\Delta_{\vec k}^2}.
\label{eq:A1}
\end{equation}
For a d-wave superconductor, $\Delta_{\vec k}=\Theta_{\vec k}\Delta_0(c_x-c_y)$.
The tunneling peak corresponds to the maximum of $E_-$ (or minimum of $E_+$).
From Fig.~\ref{fig:11}, it is sufficient to evaluate $\partial E_-/\partial c_x
$ when $c_y=\pm 1$ (the +1 solution corresponds to overdoping, -1 to 
underdoping).  In terms of the scaled variables $\tau =2t^{\prime}/t$, $\delta
=\Delta_0/2t$, and $\hat\mu =\epsilon_F/2t$, the maximum occurs at
\begin{equation}
\pm c_x={\delta^2-(1\pm\tau )(1\pm\hat\mu )\over\delta^2+(1\pm\tau )^2},
\label{eq:A2}
\end{equation}
and is $E_{max}\equiv\Delta_{TU}$ with
\begin{equation}
\bigl({E_{max}\over 2t}\bigr)^2={\delta^2(\tau +\hat\mu\pm 2)^2\over 
\delta^2+(1\pm\tau )^2},
\label{eq:A3}
\end{equation}
which give the dashed lines in Fig.~\ref{fig:12}.  The photoemission gap 
$\Delta_{PE}$ (solid line in Fig.~\ref{fig:12}) is simply the value of $E_-$ at 
$(\pi ,0)$:
\begin{equation}
\bigl({\Delta_{PE}\over 2t}\bigr)^2=(\tau -\hat\mu )^2+4\delta^2.
\label{eq:A4}
\end{equation}
When the Fermi level coincides with the VHS, $\hat\mu =\tau$, and $\Delta_{PE}=
\Delta_{TU}=2\Delta_0$, to lowest order in $\delta^2$.

\section{Properties of the CDW}

When superconductivity is absent, the Fermi surface can be found from solutions 
to
\begin{equation}
\tilde\epsilon_{\vec k}\tilde\epsilon_{\vec k+\vec Q}=G_{\vec k}^2, 
\label{eq:B1}
\end{equation}
where $\tilde\epsilon_{\vec k}=\epsilon_{\vec k}-\epsilon_F$.  When the Fermi
level is pinned at the VHS, $\epsilon_F=4t^{\prime}$, this reduces to
\begin{equation}
c_x={-(1-\tau^2)c_y\pm\sqrt{\tau^2s_y^4-g^2(1-\tau^2c_y^2)}\over 1-\tau^2c_y^2},
\label{eq:B2}
\end{equation}
using $s_y^2=1-c_y^2$ and $g=G_{\vec k}/2t$, in addition to the notation
introduced in Appendix B.  The Fermi surface shrinks to the point $(\pi /2,\pi 
/2)$ when $g=|\tau |$.

The locus of minimal gap can be found as in Appendix B.  Its exact position
depends on the orientation of the scan.  Here, we fix
$k_y$ and scan $k_x$.  The top of the $E_{k-}$ band is then at
\begin{equation}
c_x=-c_y\pm{|\tau gc_y|\over\sqrt{1-\tau^2c_y^2}}.
\label{eq:B3}
\end{equation}
In particular, for $c_y=1$
\begin{equation}
c_x=-1+{|\tau g|\over\sqrt{1-\tau^2}}.
\label{eq:B4}
\end{equation}
Note that, even though the Fermi level is pinned at the VHC, the locus of
minimum gap does not pass through $(\pi ,0)$.

{\bf $*:$} On leave of absence from Inst. of Atomic Physics, Bucharest, 
Romania

\end{document}